# Orthogonal Spin-Orbit Torque-Induced Deterministic Switching in NiO

Yixiao Qiao[1*], Zhengde Xu[1*], Zhuo Xu[1], Yumeng Yang[1,2], Zhifeng Zhu[1,2†]

[1]School of Information Science and Technology, ShanghaiTech University, Shanghai, China 201210

[2]Shanghai Engineering Research Center of Energy Efficient and Custom AI IC, Shanghai, China 201210

**Abstract**

The electrical switching of antiferromagnet (AFM) is very important for the development of ultrafast magnetic random-access memory (MRAM). This task becomes more difficult in antiferromagnetic oxide NiO which has complex anisotropy. We show that by utilizing two spin-orbit torques (SOT) from orthogonal currents, one can deterministically switch the magnetic moments of NiO in two electrical distinguishable states that can be read out using the spin Hall magnetoresistance. This deterministic switching relies on the symmetry of SOT on different sublattices, where the sign reversal of magnetic moments leads to constructive torques in the beginning and balanced torques in the end. In addition, we show that the easy-plane anisotropy plays a key role in the switching, which has been ignored in some previous works. The uniform magnetic dynamics in this work provides a clear physical picture in understanding the SOT switching of NiO. Furthermore, the electrical writing and reading function in our device advances the development of AFM-MRAM.

†Corresponding Author: zhuzhf@shanghaitech.edu.cn

Nickel oxide (NiO) is a widely used collinear antiferromagnetic (AFM) material characterized by adjacent magnetic moments aligning in an antiparallel configuration. Being an oxide, NiO provides an ideal platform for physics study,[1-4] including spin waves, spin Seebeck effects, and optical excitations. The strong internal field of AFM makes it promising in ultrafast data storage and computing applications.[5-25] However, the writing and reading of AFM still present challenges.[26,27] In the collinear AFM which has only one easy axis, previous studies have demonstrated that the damping-like torque (DLT) drives the oscillation of magnetic moments in the plane perpendicular to the spin polarization $\sigma$.[28] To achieve deterministic switching, it is predicted that one needs to have staggered effective field that act on the adjacent sublattices,[29,30] which has been experimentally demonstrated in materials like CuMnAs and $Mn_2Au$.[31] However, it requires the material to have special local crystal asymmetry, and the magnetic moments reorients in 90° instead of 180° as demonstrated in ferromagnets.[32,33] To achieve the desired 180° switching of collinear AFM, we propose to use both the uniform field-like torque (FLT) and the uniform DLT in our earlier work.[34]

Previous research have studied the biaixial NiO(001)/Pt bilayer, where the oscillatory switching is demonstrated.[35] Compared to the previous studies, in this work, we study the NiO(111) with three easy axes, which presents more complicated spin dynamics.[36] It has been proposed that a 60° switching can be achieved using a trilayer sandwich structure.[37] However, it requires that the spin currents generated by the top- and bottom-heavy metal act precisely on the two oppositely aligned magnetic moments. Other related studies involve complex domain wall dynamics.[38] Additionally, previous work has studied the role of oxygen migration in the switching behaviors, which improves our understanding about the interface effects.[39] However, achieving a uniform switching of NiO that provides a comprehensive understanding of the spin dynamics remains a challenge, which can contribute to the development of functional devices.[40-47]

In this work, we propose to use two orthogonal spin-orbit torques (SOT)[48] to deterministically switch the magnetic moments of NiO. As we will discuss later, due to the complex profile of anisotropy, the use of either one of the SOTs cannot unambiguously switch the AFM between distinguished 1 and 0 state. The use of two SOTs allows us to construct arbitrary spin polarization within the easy plane of NiO, which enables deterministic switching under a finite easy-plane anisotropy. By controlling the current direction, reversible switching can be achieved. Furthermore, we show that the two states can be easily detected using the spin Hall magnetoresistance (SMR).

The spin structure of NiO is depicted as Fig. 1(a), which consists of two AFM coupled spins in the opposite direction. The Néel vector **l** is defined as **l**=(**m**$_1$−**m**$_2$)/2. The $\varphi$ and $\theta$ are the azimuth angle and polar angle, respectively. Unlike other AFM materials, NiO has six easy axes <1 1 −2> in the {1 1 1} easy plane.[37] We assume the easy plane lies in the **x-y** plane,[37] and the easy axes are defined at $\varphi$=30°, 90°, 150°, 210°, 270°, and 330° [see the dotted lines in Fig. 1(a)]. The energy of the NiO system is described by the following Hamiltonian,[37]

$$E_{\text{NiO}} = -J_{\text{ex}}\mathbf{m}_1 \cdot \mathbf{m}_2 + K_1 \sum_i \sin^2\theta_i + K_2 \sum_i \sin^6\theta_i \cos 6\varphi_i - \mu_0 m_s \sum_i \sin^6\theta_i (\mathbf{m}_i \cdot \mathbf{H}_{\text{ext}})$$

where $i$ is the index of different sublattices. The first term is the exchange interaction with the exchange interaction constant $J_{ex} = -1.8548 \times 10^{-20}$ J.[37] The second and the third terms represent the in-plane easy-plane anisotropy and the six-fold easy-axis anisotropy with $K_1 = -6.08760 \times 10^{-24}$ J[49] and $K_2 = 1.2816 \times 10^{-26}$ J.[49] The last term represents the Zeeman interaction with the magnetic moment $m_s = 1.88\mu_B$.[37,50] The dynamics of the magnetic moments is determined by the coupled Landau-Lifshitz-Gilbert-Slonczewski (LLGS) equations:[51,52]

$$\frac{d\mathbf{m}_i}{dt} = -\gamma_i \mathbf{m}_i \times \mathbf{H}_{eff,i} + \alpha_i \mathbf{m}_i \times \frac{d\mathbf{m}_i}{dt} - \gamma_i \tau_{DLT,i} \mathbf{m}_i \times (\mathbf{m}_i \times \boldsymbol{\sigma}_i).$$

The first term on the right-hand side represents the precession of the magnetic moment around the effective magnetic field $\mathbf{H}_{eff,i} = -\frac{1}{m_s} \frac{\partial E_{NiO}}{\partial \mathbf{m}_i}$. $\gamma = 1.7609 \times 10^{11}$ rad/s·T is the gyromagnetic ratio. The second term describes the Gilbert damping with the damping constant $\alpha = 2.1 \times 10^{-5}$.[49] The last term is the SOT with the strength $\tau_{DLT} = \frac{\hbar \theta_{SH} J_c}{2eM_s t_{NiO}}$, in which the spin-Hall angle $\theta_{SH} = 0.1$, the NiO thickness $t_{NiO} = 6$ nm, the saturation magnetization $M_s = 239 \times 10^3$ A/m.[37] The coupled LLGS equations are numerically solved via the fourth order Runge-Kutta method with the time step of 5 fs.

As shown in Fig. 1(b), the device studied in this work consists of three layers, i.e., the top- and bottom-Pt layer, and the middle NiO layer. In the bottom Pt layer, we apply $\mathbf{J}_{c1}$ in the $+\mathbf{x}$ ($-\mathbf{x}$) direction, inducing a spin current that propagates upward into the NiO layer with $\boldsymbol{\sigma}_1$ in the $-\mathbf{y}$ ($+\mathbf{y}$) direction. Similarly, the current applied in the top Pt layer $\mathbf{J}_{c2}$ flows in the $+\mathbf{y}$ ($-\mathbf{y}$) direction, generating $\boldsymbol{\sigma}_2$ in the $-\mathbf{x}$ ($+\mathbf{x}$) direction. We find that small $\mathbf{J}_c$ does not alert the spin configuration of NiO. This is because the weak SOT is insufficient to overcome the energy barrier produced by the anisotropy and exchange interaction. In contrast, when a large $J_{c1} = 3\times10^{11}$ A/m$^2$ is applied, **m**$_1$ and **m**$_2$ align to the direction of **σ**$_1$, i.e., +**y** and −**y**, respectively. When $J_{c1}$ is removed, they remain in the final state because this happens to be one of the stable states. Similarly, when a large $J_{c2} = 3\times10^{11}$ A/m$^2$ is applied, **m**$_1$ and **m**$_2$ align to +**x** and −**x**, respectively. However, when $J_{c2}$ is removed, they relax to $\varphi = 30°$. Although spin reorientation can be achieved by separately applying $J_{c1}$ and $J_{c2}$, these methods are not considered reliable switching mechanisms for the following reasons. For the switching induced by $J_{c1}$, it requires precise control of easy axis along **σ**$_1$, which presents challenges in the device fabrication. In the other case, **m**$_1$ and **m**$_2$ align to the direction of **σ**$_2$, i.e., +**x** and −**x**, respectively. However, since the **x**-axis is not one of the easy axes, when $J_{c2}$ is removed, the magnetic moments align to the nearest easy axis. Since the **x**-axis lies exactly between $\varphi = 30°$ and $\varphi = 330°$, the relaxation to either of them is equally possible.

We then study the case where both $J_{c1}$ and $J_{c2}$ are applied to the trilayer structure with the initial $\varphi = 30°$. As shown in Fig. 2(a), under $J_{c1} = 1.5\times10^{11}$ A/m$^2$ and $J_{c2} = -1.5\times10^{11}$ A/m$^2$, $\varphi$ is switched from 30° (dot arrow) to 330° (solid arrow), indicating a 60° switching from one easy axis to another easy axis. The magnetization dynamics of **m**$_1$ and **m**$_2$ are shown in Figs. 2(b) and 2(c), respectively. Unlike the FM, the switching observed here mainly depends on the SOT-induced precession of the two sublattices. We find that the collaboration of the SOTs makes the two sublattices no longer collinear [see Figs. 2(d)]. This produces a large exchange field that assists the switching. To understand this switching process, we analyze the torques that are applied on the sublattices. As shown in Fig. 3(a), the DLT from $J_{c1}$ (indicated by purple arrows) acts on m1 and m2 as follows: $\boldsymbol{\tau}_{1,1} = \mathbf{m}_1 \times (\boldsymbol{\sigma}_1 \times \mathbf{m}_1)$ and $\boldsymbol{\tau}_{1,2} = \mathbf{m}_2 \times (\boldsymbol{\sigma}_1 \times \mathbf{m}_2)$. Similarly, the DLT from $\mathbf{J}_{c2}$ (indicated by green arrows) acts on m1 and m2 as follows: $\boldsymbol{\tau}_{2,1} = \mathbf{m}_1 \times (\boldsymbol{\sigma}_2 \times \mathbf{m}_1)$ and $\boldsymbol{\tau}_{2,2} = \mathbf{m}_2 \times (\boldsymbol{\sigma}_2 \times \mathbf{m}_2)$. These four torques are constructive in the beginning,[30] contributing to the reorientation of magnetization towards $\varphi = 330°$. At this state, $\boldsymbol{\tau}_{2,1}$ and $\boldsymbol{\tau}_{2,2}$ reverse direction due to the sign change of $\mathbf{m}_y$, i.e., $\mathbf{m}_{1,y} > 0$ and $\mathbf{m}_{2,y} < 0$ at $\varphi = 30°$, whereas $\mathbf{m}_{1,y} < 0$ and $\mathbf{m}_{2,y} > 0$ at $\varphi = 330°$. As a result, $\boldsymbol{\tau}_{1,1}$ is balanced with $\boldsymbol{\tau}_{2,1}$, and $\boldsymbol{\tau}_{1,2}$ is balanced with $\boldsymbol{\tau}_{2,2}$ at $\varphi = 330°$. The magnetic moments are then stabilized at this position. Using this torque analysis method, we can find that only $\mathbf{J}_{c1} > 0$ and $\mathbf{J}_{c2} < 0$ can lead to the reorientation from $\varphi = 30°$ to $330°$. When either $\mathbf{J}_{c1}$ or $\mathbf{J}_{c2}$ is reversed, the torques on $\mathbf{m}_1$ and $\mathbf{m}_2$ are in the opposite direction. Thus, they are cancelled out and the magnetic moments remain in the initial state. Similarly, when both $\mathbf{J}_{c1}$ and $\mathbf{J}_{c2}$ are reversed, it will switch from $\varphi = 30°$ to $150°$, during which $\boldsymbol{\tau}_{1,1}$ and $\boldsymbol{\tau}_{1,2}$ reverse direction due to the sign change of $\mathbf{m}_x$. Since $\mathbf{m}_1$ and $\mathbf{m}_2$ cannot be distinguished in the experiments, we consider 30° and 210°, 90° and 270°, as well as 150° and 330° to be equivalent. Therefore, the results observed here can be summarized as the first row in Table I, where only the opposite $\mathbf{J}_{c1}$ and $\mathbf{J}_{c2}$ leads to the switching from 30° to 150°. We then simulate the other cases by using different initial states. The results are summarized in Table I. It is clearly seen that the switching results are independent of the initial state, i.e., once the input stimulus is fixed, the final state is uniquely determined. This enables the use of our device in the memory application.[53]

In addition, we have studied the effect of field-like torque (FLT), which can be smaller or larger than the DLT in different systems. Therefore, different strengths of

FLT are used, i.e., 30%, 100%, and 130% of the strength of DLT. As shown in Fig. 2(b), all of them do not alter the switching process. This can also be understood by analyzing the combined FLTs from the two currents. As shown in Fig. 3(b), before the switching occurs (see the faded arrows), although the FLTs are constructive on each sublattice, the sign of FLT is opposite for different sublattices, i.e., the FLT on $\mathbf{m}_1$ and $\mathbf{m}_2$ points to $-\mathbf{z}$ and $+\mathbf{z}$, respectively. As a result, the net torque on the Neel vector vanishes, and the magnetic moments should remain in the initial state. Similarly, after the magnetization is switched (see the solid arrows), the FLTs on each sublattice compensates for each other, which will not affect the magnetization dynamics. Furthermore, previous study on NiO(001)/Pt bilayers primarily focused on biaxial anisotropy and its role in SOT-induced switching, where the biaxial anisotropy of NiO(001) originates from the formation of a four-domain structure, caused by the strain effects due to lattice mismatch between the NiO film and the SrTiO3 substrate.[35] Our work differs as we concentrate on the NiO(111) plane, which has a tri-axial symmetry, unlike the biaxial NiO(001) plane. We find that when $K_1$ is decreased by 3 times to $-2\times10^{-24}$ J, the magnetic moments develop into oscillation in the plane perpendicular to the spin polarization, and the deterministic switching cannot be achieved. This indicates that during the switching, the easy-plane anisotropy restricts the magnetic moments in the plane, which enables them to rotate to another stable state, e.g., 30° to 150°. The drastic difference observed here shows that one has to include the easy-plane anisotropy in studying the magnetic dynamics of NiO.

Finally, we note that the two states (30° and 150°) defined in this work can be easily detected in experiments using SMR. Referring to the device structure shown in Fig. 1(b), when a reading current is passed along $\varphi_{current} = 45°$ through the bottom Pt layer, the longitudinal resistivity will be different for the magnetic moment in 30° and 150° according to $\rho = \rho_0+0.5\rho_1[(1-\mathbf{m}_1\,\mathbf{t}_{read})+(1-\mathbf{m}_2\,\mathbf{t}_{read})]$, where $\mathbf{t}_{read}$ is in the **x-y** plane and orthogonal to the reading current.[54]

We show that the use of two SOTs from orthogonal currents can deterministically switch the NiO which has six-fold anisotropy. Based on the symmetry of damping-like torques, they are constructive in the beginning, and one of them changes the sign after the reorientation of the magnetic moments, resulting in a balanced torque in the end. We show that the easy-plane anisotropy is indispensable for the switching, without it the magnetic moments will evolve into oscillation. We have also verified that the switching is insensitive to the field-like torques since they are always destructive. We also propose a device structure that can be used to read out the two states based on the spin Hall magnetoresistance. The deterministic switching and the distinguishable magnetic states demonstrate the feasibility of NiO-based AFM-MRAM.

**Data Availability Statement**

The data that support the findings of this study are available from the corresponding author upon reasonable request.

**Acknowledgment**

We acknowledge the support from the National Key R&D Program of China (Grant No.

2022YFB4401700) and National Natural Science Foundation of China (Grants No. 12104301).

**Figures**

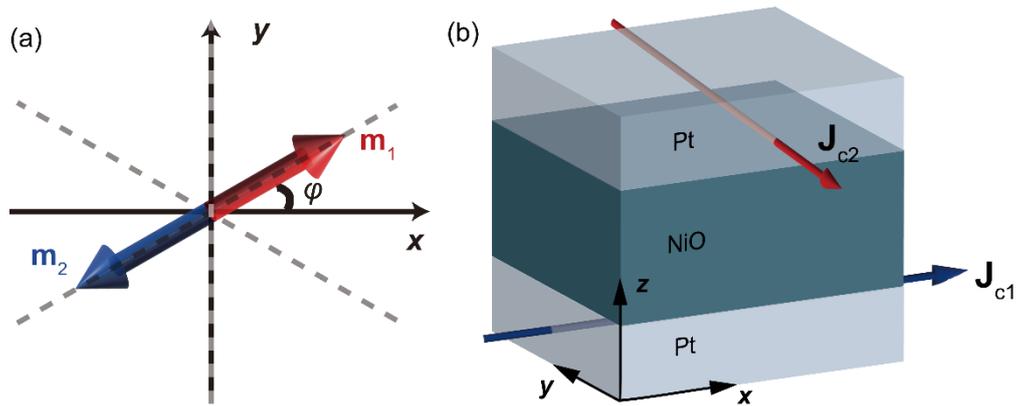

Fig. 1. (a) Schematics of the NiO system. The dashed lines denote the six-fold anisotropy. (b) Illustrations of the device structure.

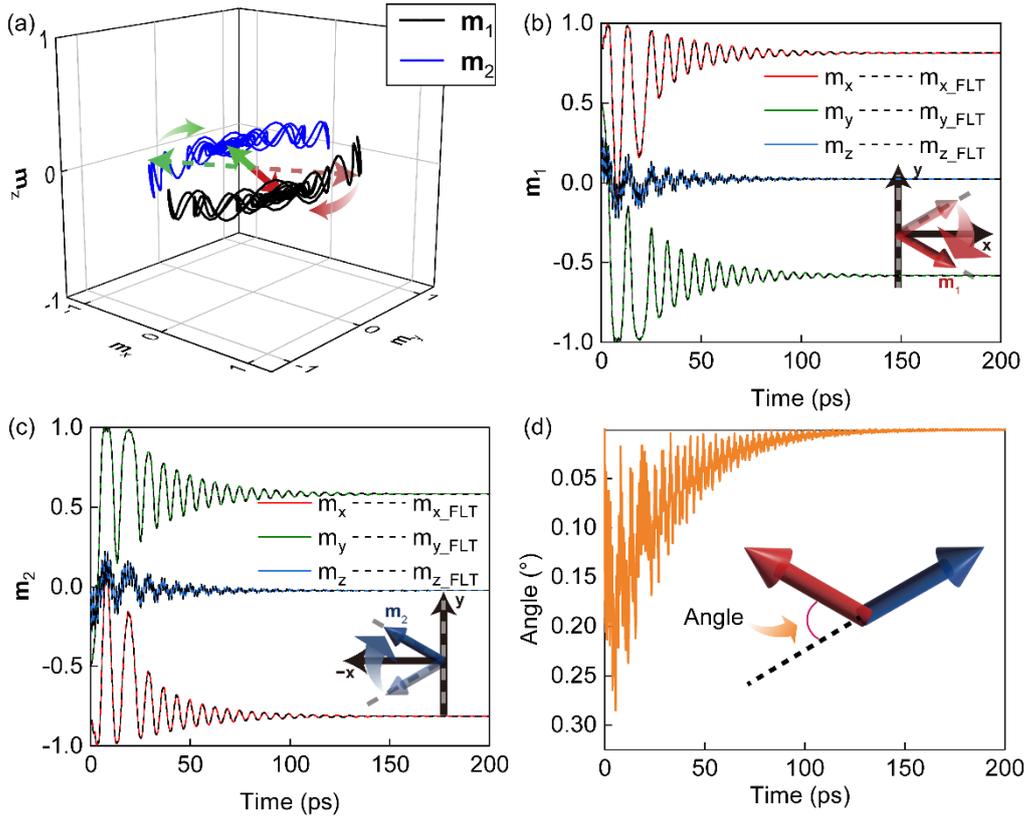

Fig. 2. (a)-(c) Switching trajectory of magnetic moments under $\mathbf{J}_{c1} = 1.5 \times 10^{11}$ A/m$^2$ and $\mathbf{J}_{c2} = -1.5 \times 10^{11}$ A/m$^2$. The strength of FLT in (b) and (c) is 30% of the DLT. (d) The angle deviation between $\mathbf{m}_1$ and $\mathbf{m}_2$.

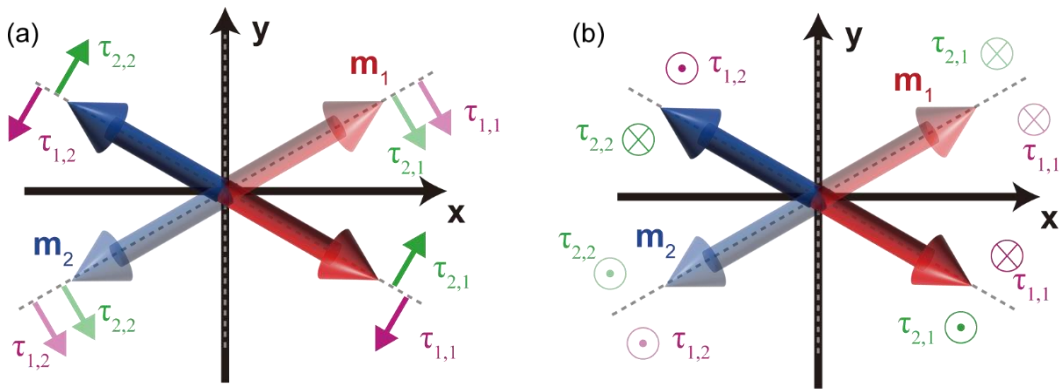

Fig. 3. Illustration of (a) DLT and (b) FLT on the magnetic moments. The faded and solid arrows denote the initial and final magnetic states, respectively.

Table I. Switching results for different initial states and current combinations

| Initial state | $J_{c1}>0, J_{c2}>0$ | $J_{c1}<0, J_{c2}<0$ | $J_{c1}>0, J_{c2}<0$ | $J_{c1}<0, J_{c2}>0$ |
|---|---|---|---|---|
| 30° | 30° | 30° | 150° | 150° |
| 90° | 30° | 30° | 150° | 150° |
| 150° | 30° | 30° | 150° | 150° |